\documentclass[aps,prb,preprint,superscriptaddress,amsmath]{revtex4}

\usepackage{dcolumn}
\usepackage{graphicx}

\begin{document}

\title{Charged impurity scattering and mobility in gated silicon nanowires}

\author{Martin P. Persson}
\affiliation{CEA-UJF, Institute for Nanosciences and Cryogenics (INAC), SP2M/L\_Sim, 17 rue des Martyrs, 38054 Grenoble Cedex 9, France}
\author{Hector Mera}
\affiliation{CEA-UJF, Institute for Nanosciences and Cryogenics (INAC), SP2M/L\_Sim, 17 rue des Martyrs, 38054 Grenoble Cedex 9, France}
\author{Yann-Michel Niquet}
\email{yniquet@cea.fr}
\affiliation{CEA-UJF, Institute for Nanosciences and Cryogenics (INAC), SP2M/L\_Sim, 17 rue des Martyrs, 38054 Grenoble Cedex 9, France}
\author{Christophe Delerue}
\email{christophe.delerue@isen.fr}
\affiliation{Institut d'Electronique, de Micro\'electronique et de Nanotechnologie (UMR CNRS 8520), D\'epartement ISEN, 41 boulevard Vauban, F-59046 Lille Cedex, France}
\author{Mamadou Diarra}
\affiliation{Institut d'Electronique, de Micro\'electronique et de Nanotechnologie (UMR CNRS 8520), D\'epartement ISEN, 41 boulevard Vauban, F-59046 Lille Cedex, France}

\begin{abstract}
We study the effects of charged impurity scattering on the electronic transport properties of $\langle 110\rangle$-oriented Si nanowires in a gate-all-around geometry, where the impurity potential is screened by the gate, gate oxide and conduction band electrons. The electronic structure of the doped nanowires is calculated with a tight-binding method and the transport properties with a Landauer-B\"uttiker Green functions approach and the linearized Boltzmann transport equation (LBTE) in the first Born approximation. Based on our numerical results we argue that: (1) There are large differences between Phosphorous- and Boron-doped systems, acceptors behaving as tunnel barriers for the electrons, while donors give rise to Fano resonances in the transmission. (2) As a consequence, the mobility is much larger in P- than in B-doped nanowires at low carrier density, but can be larger in B-doped nanowires at high carrier density. (3) The resistance of a single impurity is strongly dependent on its radial position in the nanowire, especially for acceptors. (4) As a result of subband structure and screening effects, the impurity-limited mobility can be larger in thin nanowires embedded in HfO$_2$ than in bulk Si. Acceptors might, however, strongly hinder the flow of electrons in thin nanowires embedded in SiO$_2$. (5) The perturbative LBTE largely fails to predict the correct mobilities in quantum-confined nanowires.
\end{abstract}

\maketitle

\section{Introduction}

Silicon nanowires (SiNWs) have attracted significant interest as promising building blocks for nanotechnologies. They can be fabricated by bottom-up approaches \cite{Lieber03,PYang05,Thelander05,Schmidt10} or by techniques compatible with standard complementary metal-oxide semiconductor (CMOS) process.\cite{Pei02,Colinge04,Suk05} Recently, SiNWs with diameter $d$ below 12 nm have been manufactured with excellent structural properties, which opens new opportunities for the design of nanoscale devices and for the exploration of quantum transport phenomena in low-dimensional systems. SiNWs can be used to build gate-all-around transistors in which short channel effects are reduced thanks to a better gate control,\cite{Pei02,Colinge04,Suk05,Bescond04,Lundstrom06} and transistors based on arrays of vertically stacked SiNWs with diameter close to 10 nm have been recently reported.\cite{Bera06,Dupre08,Ng09}

In this context, it is essential to understand the effects of quantum confinement on the transport properties of small SiNWs. It is well-known that the conductance would be quantized in ideal, ballistic nanowires. However, surface roughness, impurities and phonons, practically limit carrier mobilities in real devices. These scattering processes must be strongly influenced by confinement, and the physical approximations valid in bulk Si certainly break down in small enough nanowires. Recent theoretical works have therefore addressed the scattering of free carriers by phonons,\cite{Fonoberov06,Buin08} and by bulk and surface disorder in SiNWs.\cite{Markussen06,Lherbier07,Persson08} The scattering by dopants has also been studied with either density functional theory (DFT) \cite{Fernandez06,Fernandez06b,Markussen07,Rurali08} or the semi-empirical effective mass \cite{Jin07,Oh08,Martinez09,Bescond09,Bescond10} approximation. Most DFT calculations reported so far \cite{Fernandez06,Fernandez06b,Markussen07} have, however, considered \emph{neutral} dopants. Indeed, the treatment of \emph{charged} dopants within DFT is more problematic, because of the long range character of the Coulomb impurity potential, and because of intrinsic deficiencies in the present exchange-correlation functionals.\cite{Niquet10} The influence of a charged dopant on the transmission through a 2 nm diameter SiNW has nonetheless been discussed recently with DFT in Ref.~\onlinecite{Rurali08}. This work showed, in particular, that minority carriers are blocked by the impurities, but did not consider screening by the environment or free carriers, which is known to be essential from bulk\cite{Conwell50,Sanborn92}, to nanowires.\cite{Jin07} There is, therefore, a clear need for a better assessment of the effects of impurities in nanowires with more realistic potentials.

In this work, we study the scattering of electrons by \emph{charged} donor (phosphorous) and acceptor (boron) impurities in SiNWs. The scattering of electrons by acceptors occurs for example in $p$-doped transistor channels in the inversion regime. The Coulomb potential of an impurity in a nanowire is strongly dependent on its dielectric environment. In (small) free-standing nanowires, the Coulomb potential is indeed almost unscreened due to the presence of surface polarization charges in the vicinity of the impurity. Consequently, the binding energy of the dopants is much enhanced with respect to the bulk (it increases as $1/d$), which leads to a significant decrease of the doping efficiency in small ($d\lesssim 20$ nm) nanowires.\cite{Diarra07,Diarra08,Li08,Bjork09,Yoon09} In the following, we consider SiNWs surrounded by an oxide layer and a metallic gate, i.e. a \emph{gate-all-around} geometry typical of nanowire devices. In this case the impurity potential is efficiently screened by the gate, the binding energy remains close to its bulk value, and most of the donor impurities are ionized at room temperature (acceptors being usually charged negatively in the inversion regime).\cite{Diarra07,Diarra08} 

Only a few theoretical works have addressed the effect of charged impurities on the transport in gated SiNWs, with either the (perturbative) Kubo-Greenwood formula\cite{Jin07} or a (non-perturbative) Green function approach,\cite{Oh08,Martinez09,Bescond09,Bescond10} but using the effective mass approximation for the electronic structure. Our objective is to go beyond these approximations and to perform a systematic study as function of the type of impurity (donor or acceptor), its radial position in the wire, the diameter of the SiNWs and the nature of the oxide. For that purpose, we combine a tight-binding method for the electronic structure with a Landauer-B\"uttiker (LB) Green functions approach for transport. We take into account the screening of the impurity potential by the oxide, gate, and free electrons. We use the low-field mobility as a metric of the scattering strength of the impurities, which we calculate from the resistances of a representative set of single impurities. We compare Green functions with the linearized Boltzmann transport equation (LBTE) in the first Born approximation, where the impurity is treated as a perturbation.

The paper is organized as follows: We first review the methodology in section \ref{sectionMethodology}, then evidence the main trends and conclusions in the particular case of a 2 nm thick HfO$_2$ gate oxide in section \ref{sectionHfO2}. We last discuss other gate oxides and thicknesses in section \ref{sectionOtherOxides}, and summarize our conclusions in section \ref{sectionConclusion}.

\section{Methodology}
\label{sectionMethodology}

We consider cylindrical hydrogen-passivated SiNWs oriented along the $[110]$ direction. The electronic structure of the nanowires is calculated with an accurate $sp^3d^5s^*$ TB parametrization\cite{Boykin04} previously validated by {\it ab initio} calculations and comparison with experimental data.\cite{Niquet06} Each impurity is modeled by a hydrogenic potential screened by the dielectric environment, as discussed in Refs.~\onlinecite{Diarra07} and \onlinecite{Diarra08}. We assume in this respect that the SiNWs are surrounded by a conformal layer of SiO$_2$ or HfO$_2$ with thickness $t_{\rm ox}$ and a metallic gate (gate-all-around geometry). Image charge self-energy effects are included along the lines of Refs.~\onlinecite{Diarra08} and \onlinecite{Niquet06}. 

The impurity potential can also be screened by the free electrons. To account for that mechanism, we first compute the self-consistent conduction band wave functions of the homogeneous nanowire at the target carrier density $n$. We then calculate the density-density response function of the conduction band electrons with these wave functions, and solve Poisson's equation for the screened impurity potential in the linear-response approximation. This is equivalent to the so-called Random Phase Approximation (RPA) for the free carriers.\cite{Delerue05} We finally compute the impurity resistance with the Green functions approach and LBTE.

In the non-perturbative Landauer-B\"uttiker approach,\cite{Landauer57} the SiNWs are coupled to ideal semi-infinite leads and the total transmission probability $T(\varepsilon)$ is computed as a function of the electron energy $\varepsilon$ from the Green function, which is evaluated with a standard decimation technique\cite{Lherbier07,Grosso89} or a newly implemented ``knitting'' algorithm\cite{Kazymyrenko08} (for diameters $d\ge5$ nm). We consider sufficiently diluted systems and/or a generic source of incoherence (e.g., phonons) so that interference effects induced by multiple scattering events involving more than one impurity can be neglected. The resistance of a single impurity\cite{Markussen07} is then $R_{\rm imp}(\mu,T)=1/G_{\rm i}(\mu,T) - 1/G_{\rm b}(\mu,T)$, where $G_{\rm i}$ ($G_{\rm b}$) is the conductance of the nanowire with (without) impurity at temperature $T$ and chemical potential $\mu$. Both $G_{\rm i}$ and $G_{\rm b}$ are given by the finite temperature Landauer-B\"uttiker formula:
\begin{equation}
G(\mu,T)=-G_0\int d\varepsilon\,T\left(\varepsilon\right)\frac{\partial f}{\partial\varepsilon}\,,
\end{equation}
where $G_{0}=2e^2/h=(12.9{\rm\ k\Omega})^{-1}$ is the quantum of conductance (assuming spin degeneracy) and $f(\varepsilon,\mu,T)$ is the Fermi-Dirac distribution function. We will see in the following that the resistance of an impurity strongly depends on its radial position in the nanowire. We therefore define a mean resistance $\langle R_{\rm imp}\rangle$ averaged over a set of at least 16 impurity positions, from which we deduce the conductivity
\begin{equation}
\sigma=\frac{16}{\pi^2 d^4 n_{\rm i}\langle R_{\rm imp}\rangle}
\end{equation}
and the mobility $\mu=\sigma/(ne)$ (where $n$ is the free carrier density, and $n_i$ the impurity concentration).

We also compute the mobility of the SiNWs within the LBTE in the first Born approximation, treating the impurity potential as a perturbation. This approach has been widely used to calculate the carrier mobility in various materials (see for example Ref.~\onlinecite{Conwell50} for bulk Si). The relaxation time $\tau_{i}(k)$ of an electron with wavevector $k$ in subband $i$ and energy $\varepsilon_i(k)$ fulfills the following set of equations:
\begin{eqnarray}
v_{i}(k)&=&\frac{L}{\hbar}\sum_{j}\int dk'\,M_{ij}(k,k')\left[\tau_{i}(k) v_{i}(k)\right. \nonumber \\
&-& \left.\tau_{j}(k')v_{j}(k')\right]\delta\left[\varepsilon_{j}(k')-\varepsilon_{i}(k)\right]\,,
\end{eqnarray}
where $L$ is the length of the wire, $v_{i}(k)=(\partial \varepsilon_i(k)/ \partial k)/ \hbar$ is the group velocity, $j$ spans all subbands, and $M_{ij}(k,k') = |\langle j,k'|V|i,k \rangle|^2$ is a square matrix element of the impurity potential $V$. These matrix elements are computed with the unperturbed TB wave functions $|i,k \rangle$.\cite{Diarra07,Diarra08} The resistance of the impurity is then given by:
\begin{equation}
R_{\rm imp}^{-1}(\mu,T)=-\frac{e^2}{2\pi L}\sum_{i} \int dk\,\tau_{i}(k)v_{i}^2(k)\left.\frac{\partial f}{\partial\varepsilon}\right|_{\varepsilon_{i}(k)}\,.
\end{equation}
The average impurity resistance $\langle R_{\rm imp} \rangle$ and mobility are finally defined as in the Landauer-B\"uttiker approach.

\section{Case study: 2 nm thick HfO$_2$ gate oxide}
\label{sectionHfO2}

In this section, we evidence the main trends and conclusions on gate-all-around SiNWs with a 2 nm thick HfO$_2$ gate oxide. We first discuss the main features of the transmission, then the size dependence of the electron mobility, some variability issues and the screening by free carriers.

\subsection{Transmission}

\begin{figure}
\includegraphics[width = 0.66 \columnwidth]{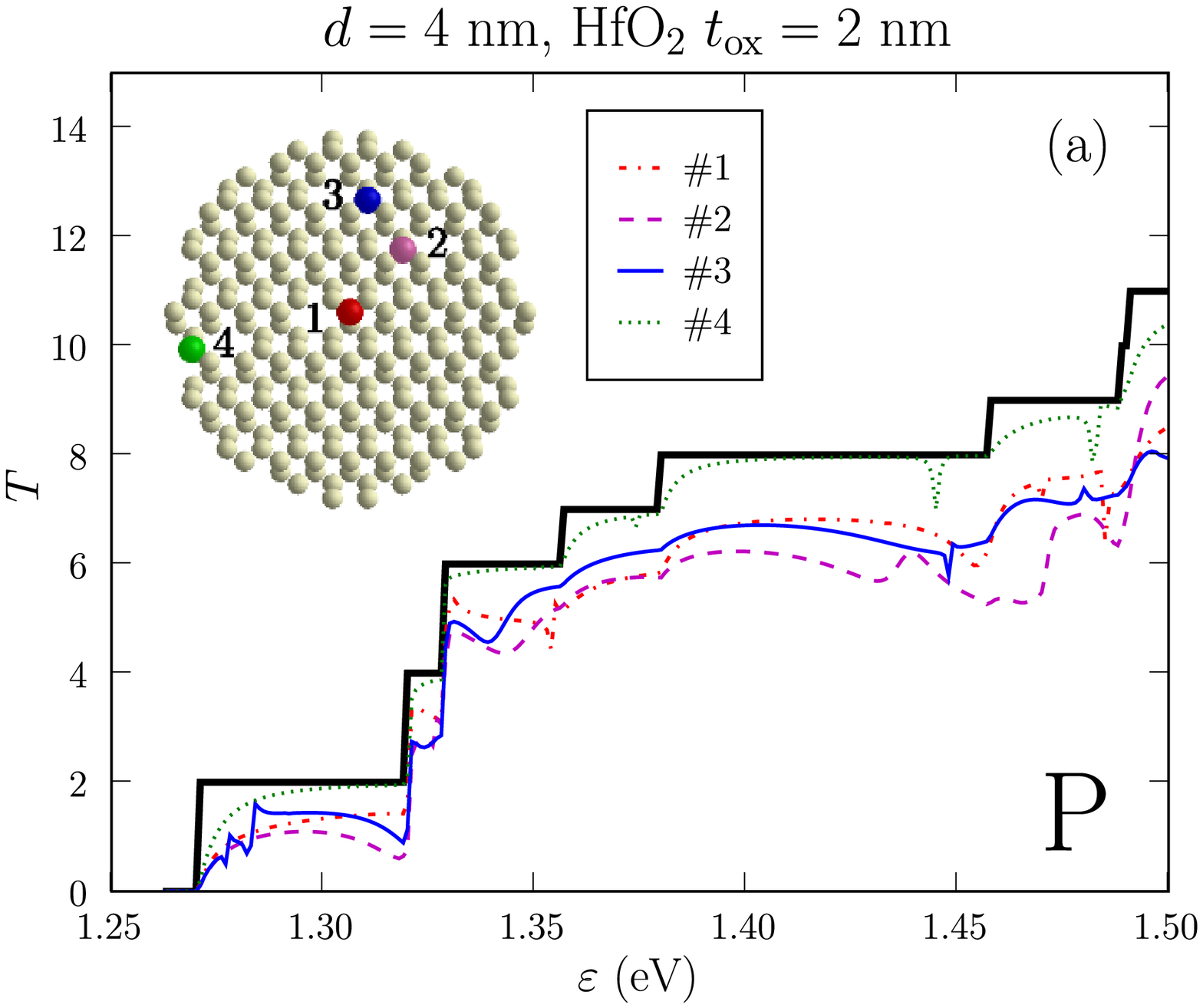}
\includegraphics[width = 0.66 \columnwidth]{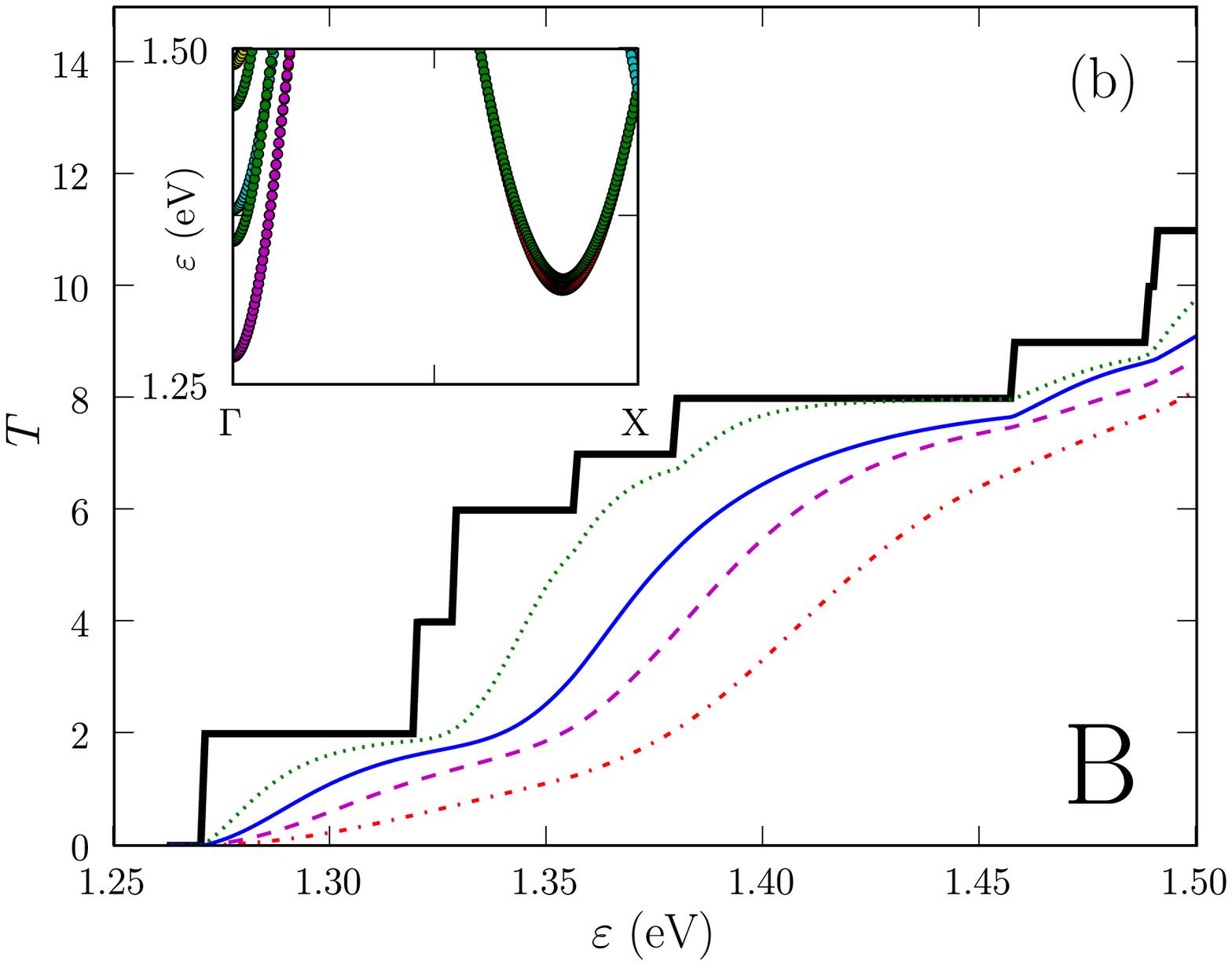}
\caption{(Color online) Total Landauer-B\"uttiker transmission $T$ as a function of the electron energy $\varepsilon$ in a 4 nm diameter SiNW with a 2 nm thick HfO$_2$ gate oxide, at low carrier concentration. The thick black line is the transmission through a pristine nanowire, while the colored lines are the transmissions through nanowires with one P (a) or one B (b) impurity. Each line corresponds to a different impurity location, shown in the inset of Fig. 1a. The band structure of the pristine nanowire is plotted in the inset of Fig. 1b.}
\label{fig_comparison_conductance}
\end{figure}

Figure \ref{fig_comparison_conductance} shows the total Landauer-B\"uttiker transmission $T(\varepsilon)$ as a function of energy, calculated for phosphorous (P) and boron (B) impurities in a 4 nm diameter SiNW. The carrier density is assumed to be low enough ($n\lesssim 10^{18}$ cm$^{-3}$, see later discussion) so that screening by free electrons can be neglected. In the absence of impurity $T(\varepsilon)$ is equal to the number of open channels (subbands) at energy $\varepsilon$; the onset of transport through a new subband gives rise (at zero temperature) to a plateau in the conductance $G=TG_0$ as a function of gate voltage. In the presence of a charged P or B dopant, the transmission is reduced due to the scattering of the electrons by the impurity potential.

Two important effects\cite{Fernandez06,Markussen07} are clearly visible in Fig. \ref{fig_comparison_conductance}: ({\it i}) The scattering strength strongly depends on the position of the impurity in the SiNW; ({\it ii}) The transmission behaves very differently for donors and acceptors. Indeed, the donor potential (Fig. \ref{fig_comparison_conductance}a) is a quantum well whose (quasi-)bound states give rise to Fano resonances, which appear as asymmetric dips and peaks in the transmission. They are typical of quantum-confined waveguides and result from the interference of the carrier wave function with the quasi-bound states of the higher-lying subbands.\cite{Tekman93,Kim99,Vargiamidis05} The number, position and width of these Fano resonances depends on the position of the impurity in the SiNW. Although they are mostly washed-out by thermal broadening at room temperature, the Fano resonances have subtle effects on the mobility, as discussed below. On the other hand acceptors behave as tunnel barriers which give rise to the smoother, resonance-free transmission curves of Fig. \ref{fig_comparison_conductance}b. The transmission is, on average, significantly smaller for B than for P impurities at low carrier density, in agreement with the above-given physical picture. The transmission in gate-all-around, B-doped nanowires remains, however, orders of magnitude larger than in free-standing SiNWs in vacuum (see Ref.~\onlinecite{Rurali08}), where the impurity is completely unscreened.\cite{noteunscreened}

\subsection{Mobility}

\begin{figure}
\includegraphics[width = 0.66 \columnwidth]{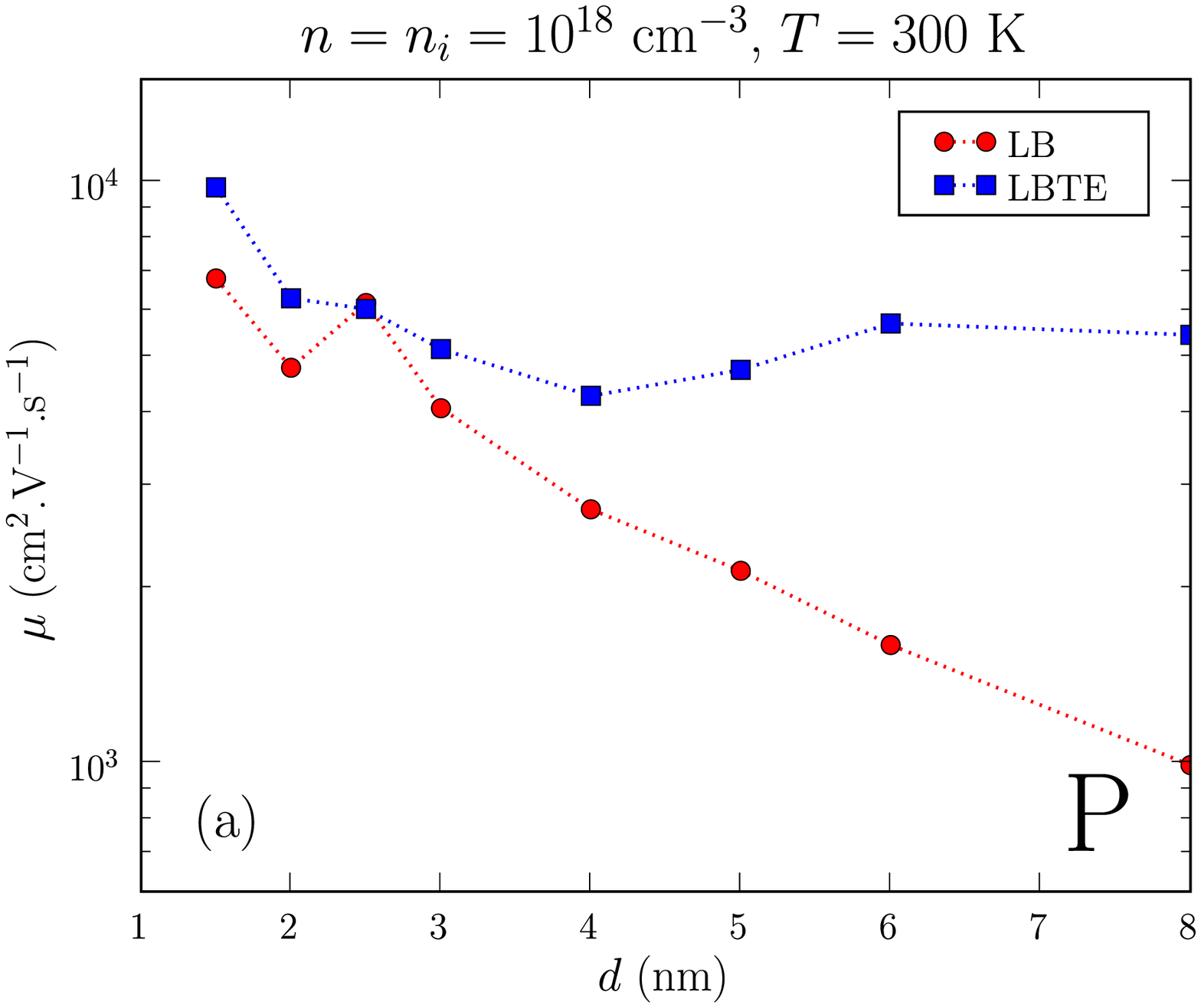}
\includegraphics[width = 0.66 \columnwidth]{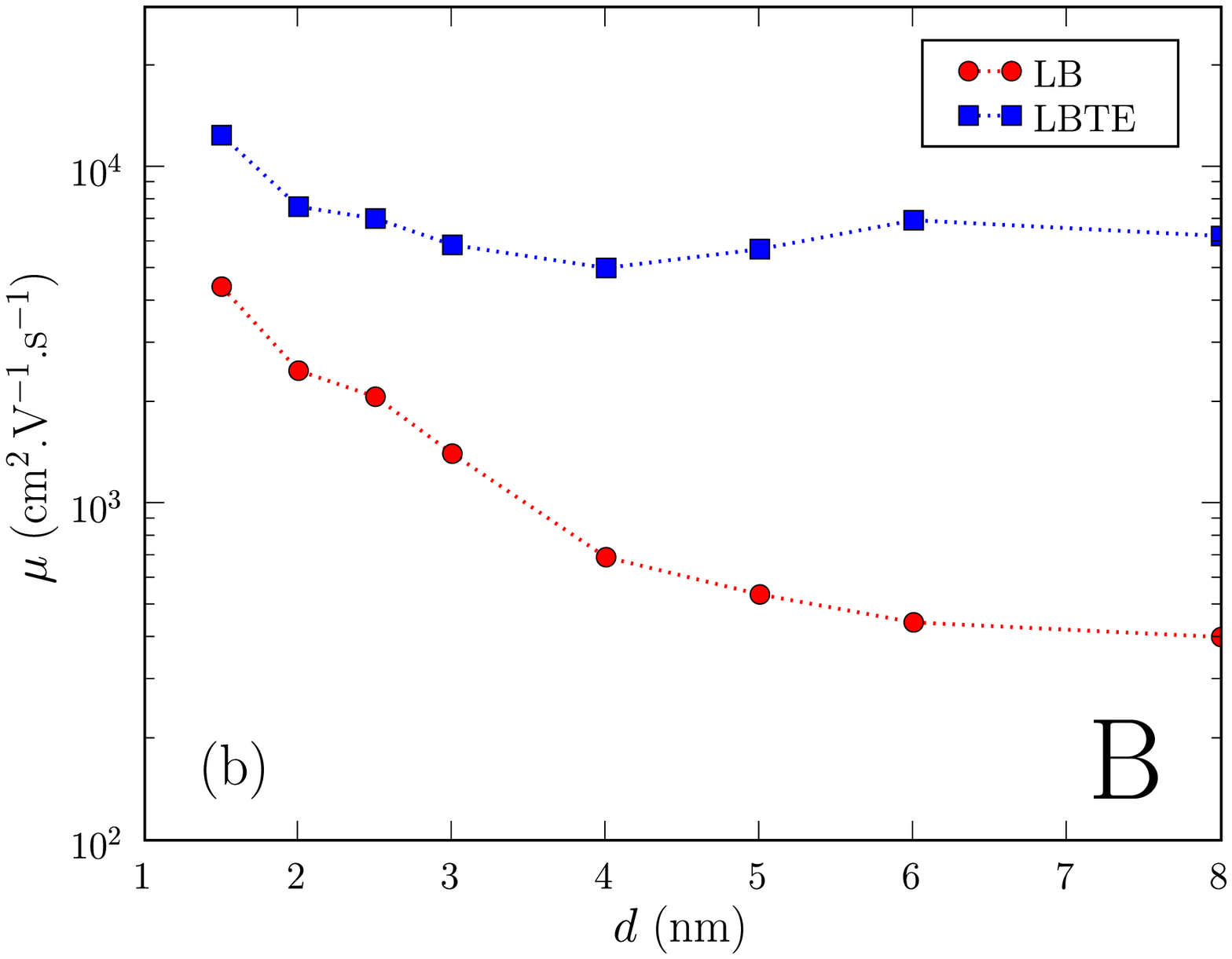}
\caption{(Color online) Room-temperature mobility as a function of the diameter of the nanowire, for P (a) and B (b) impurities (2 nm thick HfO$_2$ gate oxide, $n=n_i=10^{18}$ cm$^{-3}$). The Landauer-B\"uttiker (LB) results are compared with the linearized Boltzmann transport equation (LBTE) in the first Born approximation.}
\label{fig_mobility_diameter_comparison}
\end{figure}

The electron mobility calculated with the LB and LBTE approaches is plotted in Fig. \ref{fig_mobility_diameter_comparison} as a function of the SiNW diameter, for both donors and acceptors ($n=n_i=10^{18}$ cm$^{-3}$). As expected, the Landauer-B\"uttiker mobility in B-doped SiNWs (inversion regime) is smaller than in P-doped SiNWs because the acceptor potential acts as a barrier for the electrons. In contrast, the LBTE mobilities are -- almost\cite{note1} -- identical for P and B since the coefficients $M_{ij}(k,k')$ do not depend on the sign of the impurity potential (they are $\propto |V|^2$), a serious weakness of the first Born approximation. There is an order-of-magnitude difference between the two approaches for acceptors, and at best an order-of-magnitude agreement for donors. The error made by the LBTE is larger for acceptors because the electrons can hardly go around the barrier raised by the impurity in a nanowire, a very unfavorable situation for a perturbative approach. Although neither bound nor quasibound states can be addressed by perturbation theory, thermal broadening often helps for donors. Overall, the LBTE in the first Born approximation does not appear to be reliable enough for the prediction of impurity-limited mobilities in SiNWs, even when the potential is shorter ranged than in bulk as in gate-all-around devices.

For the HfO$_2$ gate oxide, the Landauer-B\"uttiker mobility mostly increases with decreasing wire diameter at given carrier density $n$ and impurity concentration $n_i$. This trend might appear counter-intuitive, as confinement is expected to strengthen the interaction of the carriers with the impurities. It can be explained by a combination of three factors: First, the impurities are more efficiently screened by the gate and gate oxide in small SiNWs. This is especially sensitive for acceptors, because screening reduces the height and width of the barrier the electrons have to go through. Second, confinement increases the separation between conduction subbands, which reduces the number of channels available for inter-subband scattering (see the inset of Fig. \ref{fig_comparison_conductance}b). Third, confinement also lifts the sixfold valley degeneracy of bulk Si, and splits the ground-state, twofold\cite{notevalley} degenerate $\Delta_2$ valleys at $k=0$ from the higher-lying, fourfold\cite{notevalley} degenerate $\Delta_4$ valleys at $k\ne0$.\cite{Niquet06} The $\Delta_4$ valleys therefore progressively empty with decreasing diameter, in favor of the $\Delta_2$ valleys. This enhances the mobility because the $\Delta_2$ valleys feature a lower transport mass than the $\Delta_4$ valleys.\cite{Persson08} In this respect, we would like to point out that inter-{\it valley} scattering does not significantly limit the mobility, because the range of the impurity potential, although screened by the gate and conduction band electrons, is still much larger than the unit cell (large wave vectors -- and thus very short-range potentials -- are indeed required to transfer an electron between the $\Delta_2$ and $\Delta_4$ valleys). The small fluctuations of the mobility around the main trend visible in Fig. \ref{fig_mobility_diameter_comparison} are due to band structure and Fano resonance effects.

The mobility in B-doped SiNWs tends to level slightly below $400$ cm$^2$.V$^{-1}$.s$^{-1}$ in the largest nanowires investigated in this study, while the mobility in P-doped SiNWs still shows a significant slope but bends upwards. The experimental room-temperature mobility in bulk, P- and As- doped silicon is $\mu\simeq280$ cm$^2$.V$^{-1}$.s$^{-1}$ at $n_i=10^{18}$ cm$^{-3}$, and $\mu\to1400$ cm$^2$.V$^{-1}$.s$^{-1}$ at low carrier density (phonon-limited mobility).\cite{Masetti83} Assuming that Matthiessen's rule holds and that the phonon-limited mobility is weakly dependent on the carrier density, the impurity-limited mobility in bulk $n$-type Si would therefore be $\mu_{\rm imp}\simeq350$ cm$^2$.V$^{-1}$.s$^{-1}$ at $n\simeq n_i=10^{18}$ cm$^{-3}$. More refined treatments\cite{Sanborn92,Mole99} suggest a larger $\mu_{\rm imp}\simeq650$ cm$^2$.V$^{-1}$.s$^{-1}$. The mobility of minority electrons in $p$-type Si is, of course, much less known but appears to be in the same range.\cite{Swirhun86} These data imply that ({\it i}) the impurity-limited mobility can be larger in thin gate-all-around devices than in bulk silicon, and ({\it ii}) that the impurity-limited mobility in B-doped SiNWs embedded in HfO$_2$ might exhibit a shallow minimum in the $d>10$ nm range. We will further discuss these issues for different gate oxides and thicknesses in section \ref{sectionOtherOxides}.

\subsection{Variability}

\begin{figure}
\includegraphics[width = 0.66 \columnwidth]{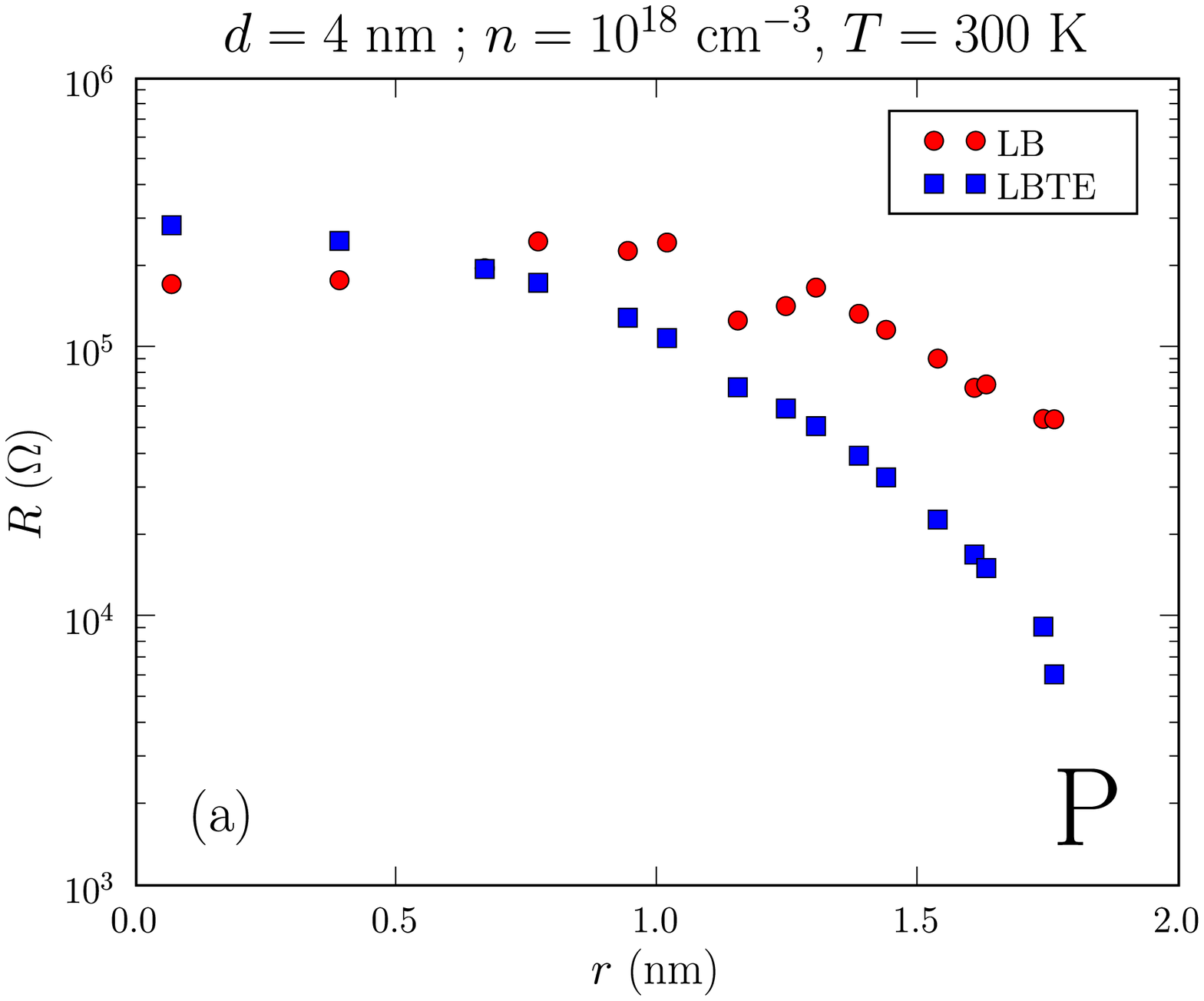}
\includegraphics[width = 0.66 \columnwidth]{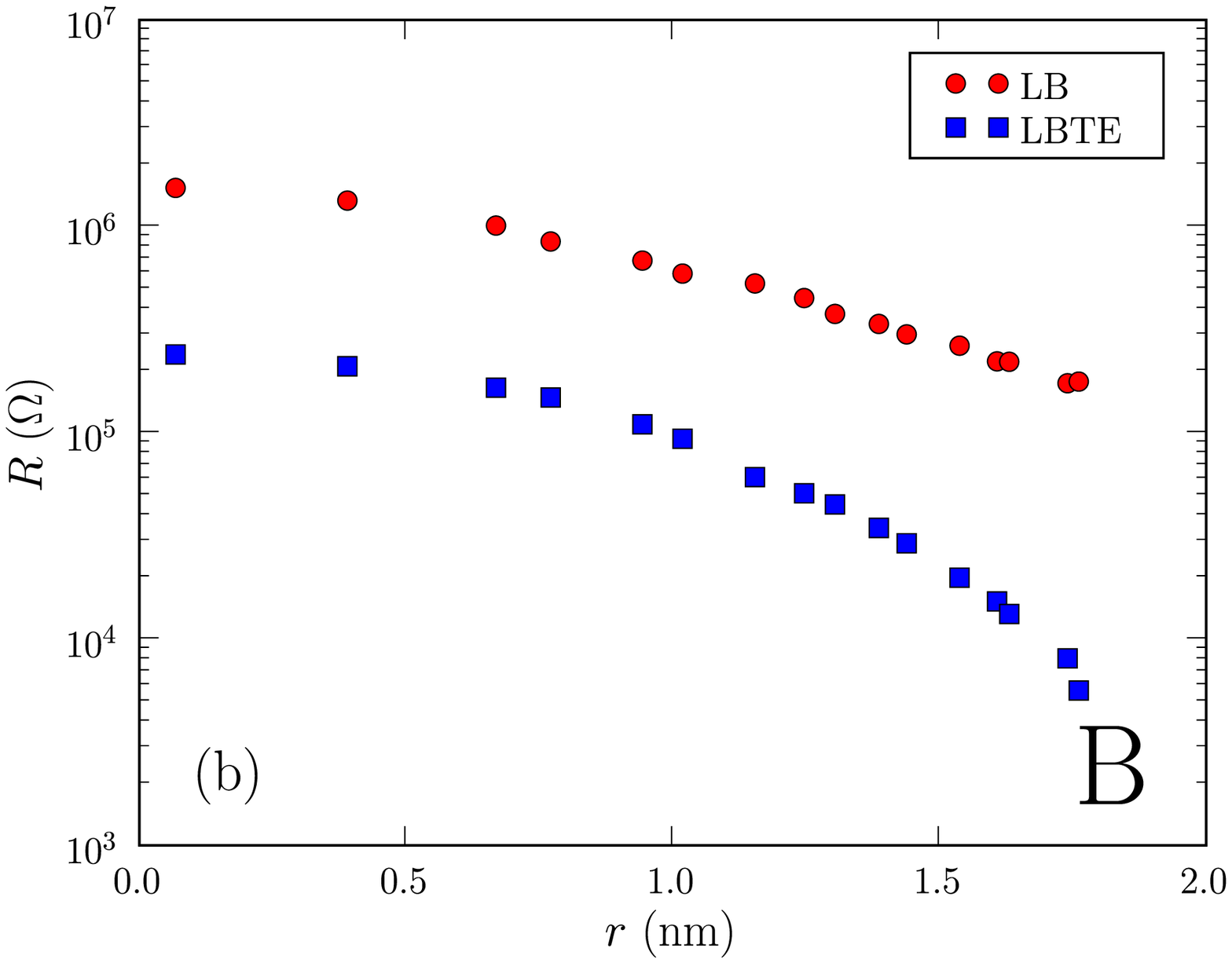}
\caption{(Color online) Resistance of a single P (a) or B (b) impurity in a 4 nm diameter SiNW as a function of its radial position (2 nm thick HfO$_2$ gate oxide, $n=10^{18}$ cm$^{-3}$). The Landauer-B\"uttiker (LB) results are compared with the linearized Boltzmann transport equation (LBTE) in the first Born approximation.}
\label{fig_resistance_radial}
\end{figure}

The resistance of single impurities in a 4 nm diameter SiNW is plotted as a function of their radial coordinate in Fig. \ref{fig_resistance_radial}, in both the LB and LBTE approaches ($n=10^{18}$ cm$^{-3}$). As expected from Fig. \ref{fig_comparison_conductance}, the resistance of an impurity is strongly dependent on its radial position in the nanowire (the angular dependence being much weaker). It tends to decrease close to the surface as the impurity moves out of the main flow of electrons and is better screened by the gate. While the random fluctuation of the number of dopants in ultimate transistors is already considered as a major issue in the microelectronics industry,\cite{Kuhn08} our results show that the fluctuation of the impurity positions also contributes to the variability in thin SiNW transistors. The resistance of single B impurities is monotonously decreasing from the center to the surface of the SiNW. It spans around one order of magnitude (for diameters $d>2$ nm), due to the sensitivity of the tunneling current to the barrier height and width. The resistance of single P impurities is more weakly dependent on their radial position and might be non-monotonous. For example, the sharp feature around $r=1$ nm in Fig. \ref{fig_resistance_radial}a coincides with a rapid change in the distribution of Fano resonances in the first subband [see the magenta (\#2) and blue (\#3) curves in Fig. \ref{fig_comparison_conductance}]. For both P and B, the difference between the LB and LBTE resistances is maximum at the surface, because the matrix elements of the potential, $M_{ij}(k, k')$, decrease too fast in the first Born approximation when the impurity moves out of the electron flow.

\subsection{Screening by free carriers}

\begin{figure}
\includegraphics[width = 0.66 \columnwidth]{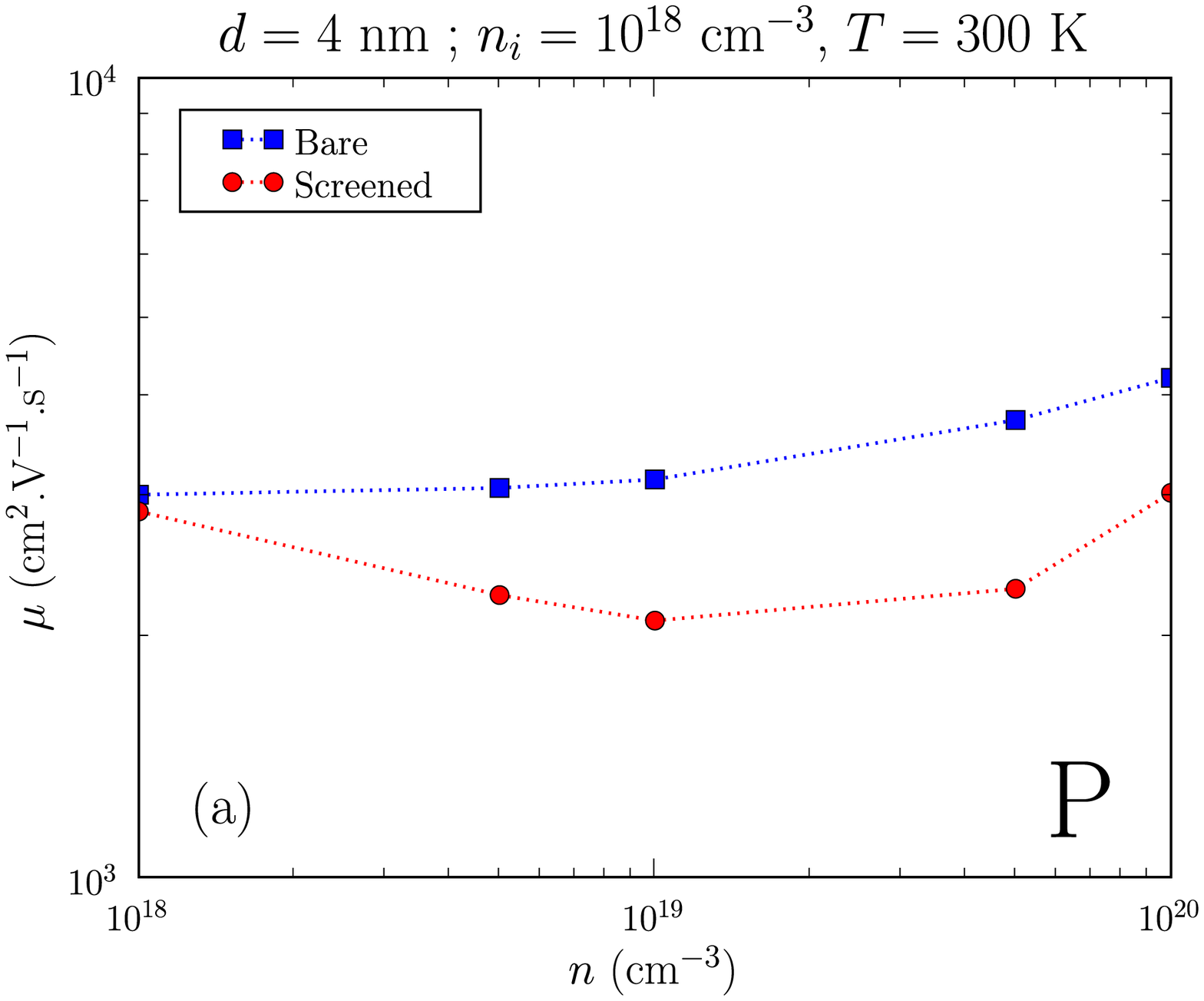}
\includegraphics[width = 0.66 \columnwidth]{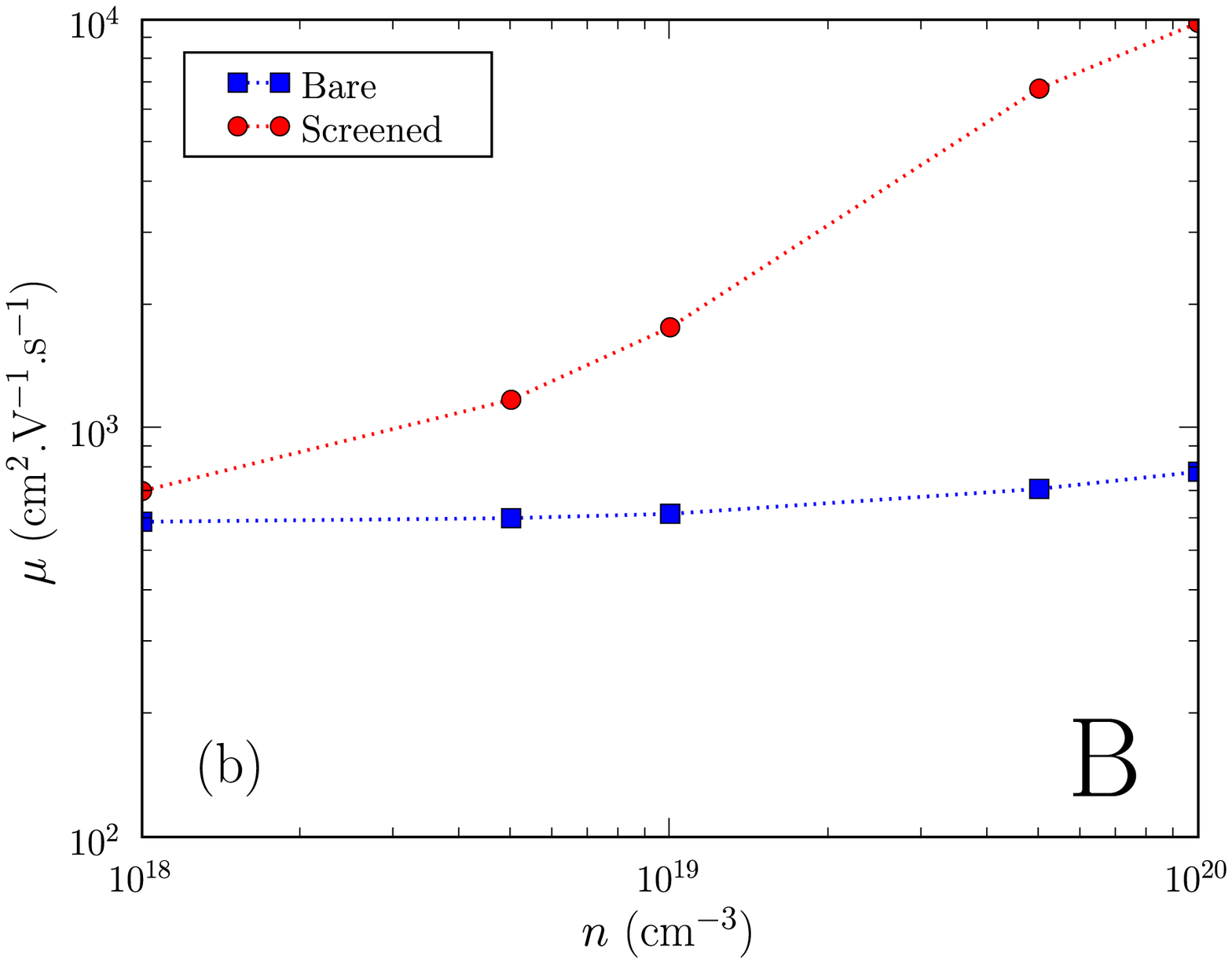}
\caption{(Color online) Landauer-B\"uttiker mobility in a P-doped (a) and B-doped (b) 4 nm diameter SiNW as function of the conduction band electron density $n$ (2 nm thick HfO$_2$ gate oxide, $n_i=10^{18}$ cm$^{-3}$). In the ``bare'' case only the screening by the gate and gate oxide is taken into account. For Boron, screening by charge carriers reduces the range of the impurity potential, which increases the mobility. The counter-intuitive behavior of the mobility in P-doped nanowires results from the interplay between the impurity well and lateral quantum confinement (see text).}
\label{fig_density}
\end{figure}

In practice, the carrier density $n$ can be modulated by the gate voltage. The Landauer-B\"uttiker mobility in a 4 nm diameter SiNW is plotted as a function of $n$ in Fig. \ref{fig_density}, for both donors and acceptors with a concentration $n_i=10^{18}$ cm$^{-3}$. The ``bare'' data does not take screening by the free electrons into account, while the ``screened'' data does. The trends evidenced in Fig. \ref{fig_density} are typical of the investigated diameter range.

At room temperature, the bare mobility is essentially constant in the whole $n<10^{20}$ cm$^{-3}$ range, which shows that mobility is a relevant concept in long channels down to the smallest SiNWs. The bare and screened mobilities almost coincide at low carrier densities $n\lesssim 10^{18}$ cm$^{-3}$ where the impurities are mainly screened by the gate and gate oxide only. As the electron density is further increased, the free carriers effectively reduce the range and depth of the impurity potential. This strongly enhances, as expected, the mobility in B-doped SiNWs (by around one order of magnitude at $n=10^{20}$ cm$^{-3}$). The mobility in P-doped nanowires remains, however, almost constant up to $n=10^{20}$ cm$^{-3}$. It even features a shallow minimum around $n=10^{19}$ cm$^{-3}$, which means that the conductivity increases sub-linearly with carrier density. This counter-intuitive trend results from the interplay between the impurity well and lateral quantum confinement, and from the complex behavior of Fano resonances. Indeed, as shown for example in Figs. 1 and 2 and in the appendix of Ref.~\onlinecite{Vargiamidis05}, the decrease of the depth (or width) of a well placed along a quantum-confined electron waveguide does not necessarily improve the background transmission through this waveguide. In addition, the Fano resonances of the first subband are pushed closer the edge of the second subband as the range of the potential decreases, which markdely affects the transmission profile around the Fermi energy. As a consequence, the mobility in B-doped nanowires can be larger than the mobility in P-doped nanowires at ``high'' carrier density $n\gtrsim 10^{19}$ cm$^{-3}$. This again shows that screening -- either by the dielectric environment or by the free carriers -- can not be neglected when discussing the transport properties of SiNWs in the inversion regime.

\section{Role of the gate oxide}
\label{sectionOtherOxides}

In this section, we discuss the mobility in P- and B- doped SiNWs with different gate oxides and thicknesses.

\begin{figure}
\includegraphics[width = 0.66 \columnwidth]{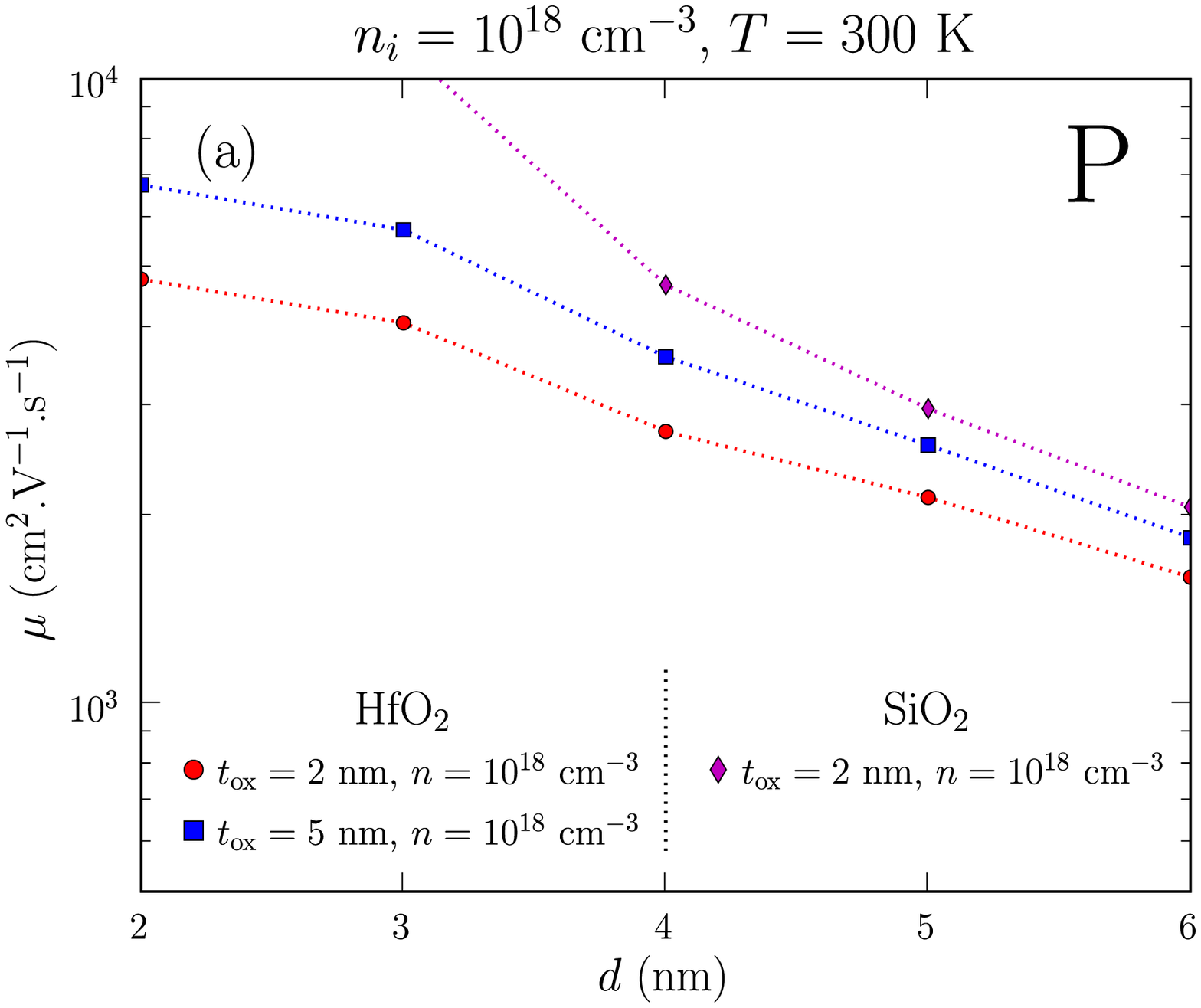}
\includegraphics[width = 0.66 \columnwidth]{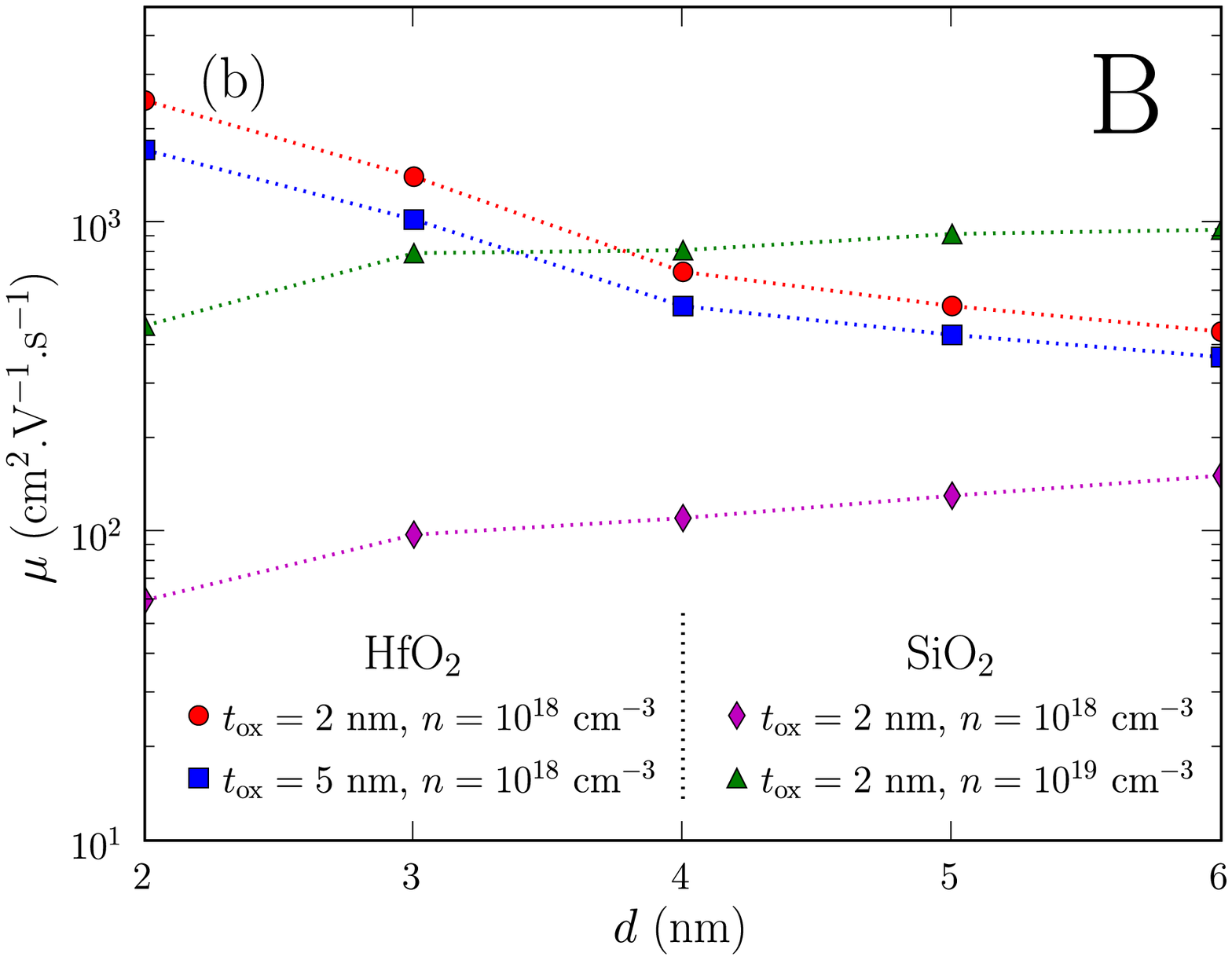}
\caption{(Color online) Room-temperature mobility as a function of the diameter of the nanowire, for P (a) and B (b) impurities, and for different gate oxides, oxide thicknesses and carrier densities.}
\label{figOxides}
\end{figure}

The room-temperature mobility in P- and B- doped SiNWs is plotted as a function of the nanowire diameter in Fig. \ref{figOxides}, for a 2 nm and a 5 nm thick HfO$_2$ gate oxide, as well as for a 2 nm thick SiO$_2$ gate oxide. The impurity concentration is $n_i=10^{18}$ cm$^{-3}$; The carrier density is $n=10^{18}$ cm$^{-3}$ or $n=10^{19}$ cm$^{-3}$. The impurity potentials are more shallow in HfO$_2$ than in SiO$_2$ (due to the larger dielectric constant), but get even shorter-ranged when the thickness of the oxide decreases (the range of the potential is, indeed, roughly proportional to the gate radius at low carrier densities).

The mobility in SiNWs embedded in HfO$_2$ is weakly dependent on the oxide thickness. Indeed, a few nanometers of such a high-$\kappa$ material are enough to screen the impurities almost completely, so that the effect of the gate becomes insignificant. Still, as expected, the mobility in B-doped SiNWs increases when decreasing the oxide thickness. However, the mobility in P-doped SiNWs slightly decreases with decreasing $t_{\rm ox}$, again showing that a better screening does not necessarily come with an enhancement of the mobility in P-doped quantum-confined nanowires.

\begin{figure}
\includegraphics[width = 0.66 \columnwidth]{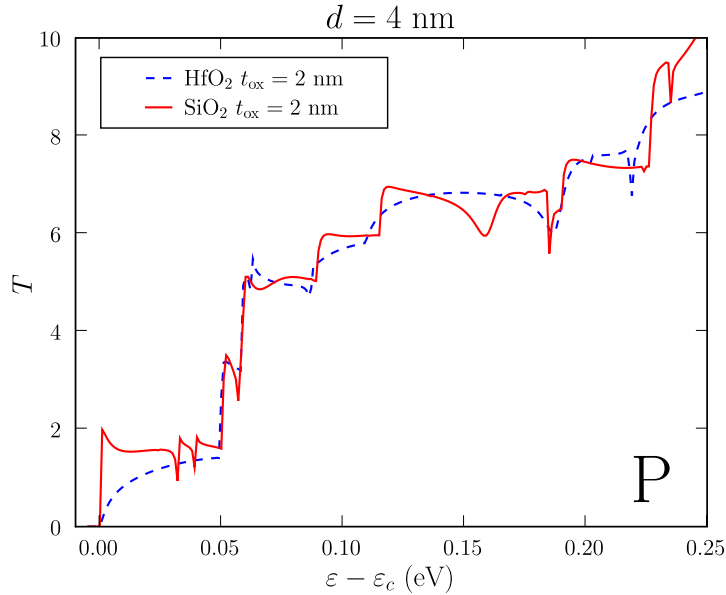}
\caption{(Color online) Total Landauer-B\"uttiker transmission $T$ as a function of the electron energy in a 4 nm diameter SiNW with a 2 nm thick SiO$_2$ or HfO$_2$ gate oxide, for the P impurity closest to the axis of the nanowire. The carrier density is $n=10^{18}$ cm$^{-3}$. The electron energy $\varepsilon$ is measured with respect to the band edge $\varepsilon_c$, which is different in SiO$_2$ and HfO$_2$ due to image charge self-energy\cite{Niquet06} and self-consistency effects.}
\label{figToxP}
\end{figure}

This is further evidenced by the SiO$_2$ data. Again, the mobility in P-doped SiNWs increases (with respect to HfO$_2$) despite the lower dielectric constant. The potential landscape around the impurity is indeed very different in SiO$_2$ and HfO$_2$ (see later discussion for B impurities), which leads to different background transmission profiles and Fano resonances. This is illustrated in Fig. \ref{figToxP} for a 4 nm diameter SiNW: while the transmission is small around the conduction band edge in HfO$_2$, it is finite in SiO$_2$, and shows more and stronger resonances in the first subbband. Still, the difference between SiO$_2$- and HfO$_2$-coated P-doped SiNWs decreases with increasing nanowire diameter as quantum confinement gets weaker and the transport becomes multi-band.

\begin{figure}
\includegraphics[width = 0.66 \columnwidth]{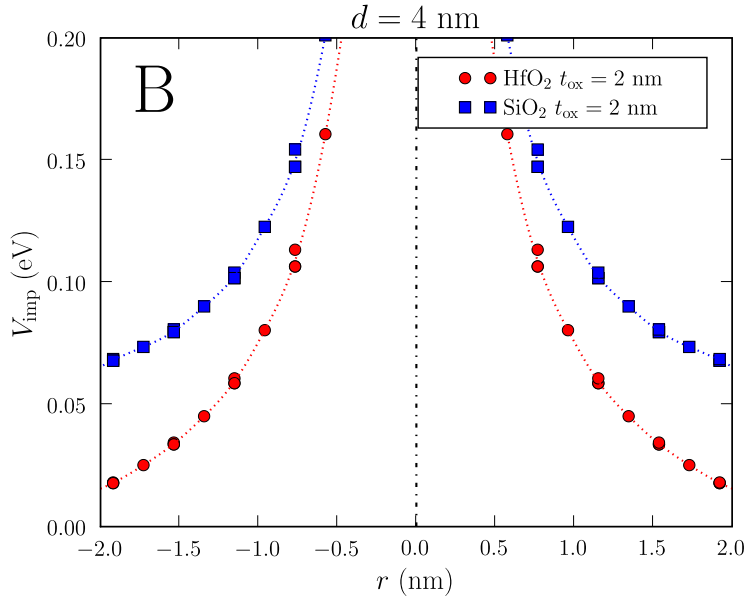}
\caption{(Color online) The impurity potential $V_{\rm imp}$ created by the B impurity closest to the axis of a 4 nm diameter SiNW with a 2 nm thick SiO$_2$ or HfO$_2$ gate oxide. The potential is plotted on selected atoms along a $[1\bar10]$ axis perpendicular to the nanowire. The dotted lines are just guides to the eyes. The carrier density is $n=10^{18}$ cm$^{-3}$. The potential is much larger at the surface of the nanowire in SiO$_2$ than in HfO$_2$, hindering the flow of carriers throughout the whole cross section of the SiNW.}
\label{figVB}
\end{figure}

The mobility in B-doped SiNWs embedded in SiO$_2$ is strongly hindered at low carrier densities. At variance with all previous cases, it increases with increasing nanowire diameter. The mobility in these nanowires is actually limited by the lateral extension of the barrier raised by the acceptor. This is illustrated in Fig. \ref{figVB}, which represents the potential of a B impurity in the cross section of a 4 nm diameter nanowire. In HfO$_2$, the impurity potential is almost zero already at the surface of the nanowire, allowing for significant transmission even at low carrier energy. In SiO$_2$, the potential is still sizeable at the surface of the nanowire, effectively preventing the flow of carriers throughout the whole cross section over $\simeq 50$ meV above the conduction band edge, and quenching the mobility. As the nanowire diameter increases, the height of this residual barrier decreases (because there is more space around the impurity for the potential to decay), and the mobility increases. Also, this barrier can be significantly lowered by the free carriers, as evidenced by the $n=10^{19}$ cm$^{-3}$ data in Fig. \ref{figOxides}. Although this might increase the $I_{\rm on}/I_{\rm off}$ ratio of B-doped channels embedded in SiO$_2$, we conclude from these results that the introduction of high-$\kappa$ oxides is mandatory in ultimate nanowire devices to prevent strong impurity scattering.

\section{Conclusions}
\label{sectionConclusion}

In conclusion, the effects of charged (P and B) impurities on the electron transport properties of $\langle 110\rangle$-oriented gate-all-around SiNWs with diameters $d\le8$ nm have been analyzed using the Landauer-B\"uttiker approach and the linearized Boltzmann transport equation (LBTE). The main results of our calculations are: 

(1) Whereas the transmission in B-doped SiNWs presents a monotonous dependence on energy, the transmission in P-doped SiNWs shows a complex behavior with multiple Fano resonances. 

(2) At low carrier density, the mobility is much larger in P-doped nanowires in accumulation than in B-doped nanowires in inversion. 

(3) The resistance of single impurities (especially acceptors) strongly depends on their radial position in the nanowire, which may represent an important source of variability in ultimate transistors based on SiNWs. 

(4) The mobility in P-doped quantum confined SiNWs does not necessarily increase when the impurities are better screened. In particular, the mobility is weakly dependent on the carrier density in P-doped SiNWs embedded in HfO$_2$, while it rapidly increases with carrier density in B-doped SiNWs. As a consequence, the mobility can be larger in B- than in P-doped SiNWs at high carrier density. 

(5) In SiNWs embedded in HfO$_2$, the impurity-limited mobility increases with decreasing wire diameter, and can be larger in the smallest nanowires than in bulk Si. On the opposite, acceptors might severely quench the mobility in B-doped SiNWs embedded in SiO$_2$. The use of high-$\kappa$ gate oxides is therefore mandatory in ultimate SiNW devices to prevent strong impurity scattering. 

(6) The error made by the perturbative LBTE with respect to the Landauer-B\"uttiker approach is usually very large in quantum-confined nanowires.

\section*{Acknowledgements}

This work was supported by the EU Project No. 015783 NODE, by the French National Research Agency (ANR) projects PREEANS and QUANTAMONDE. The calculations were performed at the CCRT and CINES supercomputing centers.


\begin{thebibliography}{99}

%General paper on nanowires
\bibitem{Lieber03} C.~M. Lieber, Mater. Res. Soc. Bull. {\bf 28}, 486 (2003).
\bibitem{PYang05} P. Yang, Mater. Res. Soc. Bull. {\bf 30}, 85 (2005).
\bibitem{Thelander05} C. Thelander, H.~A. Nilsson, L.~E. Jensen, and L. Samuelson, Nano Lett. {\bf 5}, 635 (2005).
\bibitem{Schmidt10} V. Schmidt, J.~V. Wittemann, and U. G\"osele, Chem. Rev. {\bf 110}, 361 (2010).

%Gate all-around & transistors
\bibitem{Pei02} G. Pei, J. Kedsierski, P. Oldiges, M. Ieong, and E.~C.~C. Kan, IEEE Trans. Electron Devices {\bf 49}, 1411 (2002).
\bibitem{Colinge04} P. Colinge, Solid-State Electron. {\bf 48}, 897 (2004).
\bibitem{Suk05} S.~D. Suk, S.-Y. Lee, S.-M. Kim, E.-J. Yoon, M.-S. Kim, M. Li, C.~W. Oh, K.~H. Yeo, S.~H. Kim, D.-S. Shin, K.-H. Lee, H.~S. Park, J.~N. Han, C.~J. Park, J.-B. Park, D.-W. Kim, D. Park, and B.-I. Ryu, IEDM Tech. Digest 2005, 717 (2005).
\bibitem{Bescond04} M. Bescond, J.~L. Autran, D. Munteanu, and M. Lannoo, Solid-State Electron. {\bf 48}, 567 (2004).
\bibitem{Lundstrom06} M. Lundstrom and J. Guo, {\it Nanoscale Transistors: Device Physics, Modeling and Simulation} (Springer, 2006).

%vertically stacked transistors
\bibitem{Bera06} L.~K. Bera, H.~S. Nguyen, N. Singh, T.~Y. Liow, D.~X. Huang, K.~M. Hoe, C.~H. Tung, W.~W. Fang, S.~C. Rustagi, Y. Jiang, G.~Q. Lo, N. Balasubramanian, and D.~L. Kwong, IEDM Tech. Digest 2006, 551 (2006).
\bibitem{Dupre08} C. Dupr\'e, A. Hubert, S. B\'ecu, M. Jublot, V. Maffini-Alvaro, C. Vizioz, F. Aussenac, C. Arvet, S. Barnola, J.-M. Hartmann, G. Garnier, F. Allain, J.-P. Colonna, M. Rivoire, L. Baud, S. Pauliac, V. Loup, P. Rivallin, B. Guillaumot, G. Ghibaudo, O. Faynot, T. Ernst, and S. Deleonibus, IEDM Tech. Digest 2008, 749 (2008).
\bibitem{Ng09} R.~M.~Y. Ng, T. Wang, F. Liu, X. Zuo, J. He, and M. Chan, IEEE Electron Device Lett. {\bf 30}, 520 (2009).

%phonons
\bibitem{Fonoberov06} V.~A. Fonoberov and A.~A. Balandin, Nano Lett. {\bf 6}, 2442 (2006).
\bibitem{Buin08} K. Buin, A. Verma, A. Svizhenko, and M.~P. Anantram, Nano Lett. {\bf 8}, 760 (2008).

%bulk and surface disorder
\bibitem{Markussen06} T. Markussen, R. Rurali, M. Brandbyge, and A.-P. Jauho, Phys. Rev. B {\bf 74}, 245313 (2006).
\bibitem{Lherbier07} A. Lherbier, M.~P. Persson, Y.~M. Niquet, F. Triozon, and S. Roche, Phys. Rev. B {\bf 77}, 085301 (2008).
\bibitem{Persson08} M.~P. Persson, A. Lherbier, Y.~M. Niquet, F. Triozon, and S. Roche, Nano Letters {\bf 8}, 4146 (2008).

%ab initio dopants
\bibitem{Fernandez06} M.~V. Fern\'andez-Serra, Ch. Adessi, and X. Blase, Nano Lett. {\bf 6}, 2674 (2006).
\bibitem{Fernandez06b} M.~V. Fern\'andez-Serra, Ch. Adessi, and X. Blase, Phys. Rev. Lett. {\bf 96}, 166805 (2006).
\bibitem{Markussen07} T. Markussen, R. Rurali, A.-P. Jauho, and M. Brandbyge, Phys. Rev. Lett. {\bf 99}, 076803 (2007).
\bibitem{Rurali08} R. Rurali, T. Markussen, M. Brandbyge, and A.-P. Jauho, Nano Lett. {\bf 8}, 2825 (2008).
%\bibitem{Rurali09} R. Rurali, B. Aradi, T. Frauenheim, and A. Gali, Phys. Rev. B {\bf 79}, 115303 (2009).

%theory ionized-impurity scattering in gate all around transistors
\bibitem{Jin07} S. Jin, M.~V. Fischetti, and T.-W. Tang, J. Appl. Phys. {\bf 102}, 083715 (2007).
\bibitem{Oh08} J.~H. Oh, D. Ahn, Y.~S. Yu, and S.~W. Hwang, Phys. Rev. B {\bf 77}, 035313 (2008).
\bibitem{Martinez09} A. Martinez, A.~R. Brown, N. Seoane, and A. Asenov, Journal of Physics: Conference Series {\bf 193}, 012047 (2009).
\bibitem{Bescond09} M. Bescond, M. Lannoo, F. Michelini, L. Raymond, and M.~G. Pala, Microelectronics Journal {\bf 40}, 756 (2009).
\bibitem{Bescond10} M. Bescond, M. Lannoo, L. Raymond and F. Michelini, J. Appl. Phys. {\bf 107}, 093703 (2010).

%paper DFT deficiencies  Y.~M. Niquet, L. Genovese, C. Delerue and T. Deutsch, Phys. Rev. B {\bf 81}, 161301(R) (2010) 
\bibitem{Niquet10}  Y.~M. Niquet, L. Genovese, C. Delerue and T. Deutsch, Phys. Rev. B {\bf 81}, 161301(R) (2010) 

%Theory perturbation bulk
\bibitem{Conwell50} E.~M. Conwell and V.~F. Weisskopf, Phys. Rev. {\bf 77}, 388 (1950); H. Brooks and C. Herring, Phys. Rev. {\bf 83}, 879 (1951).

% Mobility in bulk.
\bibitem{Sanborn92} B.~A. Sanborn, P.~B. Allen, and G.~D. Mahan, Phys. Rev. B {\bf 46}, 15123 (1992).

%Theory dopants nanowires
\bibitem{Diarra07} M. Diarra, Y.~M. Niquet, C. Delerue, and G. Allan, Phys. Rev. B {\bf 75}, 045301 (2007).
\bibitem{Diarra08} M. Diarra, C. Delerue, Y.~M. Niquet, and G. Allan, J. Appl. Phys. {\bf 103}, 073703 (2008).
\bibitem{Li08} B. Li, A.~F. Slachmuylders, B. Partoens, W. Magnus, and F.~M. Peeters, Phys. Rev. B {\bf 77}, 115335 (2008).

%Exp dopants
\bibitem{Bjork09} M.~T. Bj\"ork, H. Schmid, J. Knoch, H. Riel, and W. Riess, Nature Nanotechnology {\bf 4}, 103 (2009).
\bibitem{Yoon09} J. Yoon, A.~M. Girgis, I. Shalish, L.~R. Ram-Mohan, and V. Narayanamurti, Appl. Phys. Lett. {\bf 94}, 142102 (2009).

%Tight binding
\bibitem{Boykin04} T.~B. Boykin, G. Klimeck, and F. Oyafuso, Phys. Rev. B {\bf 69}, 115201 (2004).
\bibitem{Niquet06} Y.~M. Niquet, A. Lherbier, N.~H. Quang, M.~V. Fern\'andez-Serra, X. Blase, and C. Delerue, Phys. Rev. B {\bf 73}, 165319 (2006).
\bibitem{Delerue05} C. Delerue and M. Lannoo, {\it Nanostructures: Theory and Modelling} (Springer, 2004).

%Landauer
\bibitem{Landauer57} R. Landauer, IBM J. Res. Dev. {\bf 1}, 223 (1957); M. B\"uttiker and R. Landauer, Phys. Rev. Lett. {\bf 49}, 1739 (1982).
\bibitem{Grosso89} G. Grosso, S. Moroni, and G.~P Parravicini, Phys. Rev. B {\bf 40}, 12328 (1989).
\bibitem{Kazymyrenko08} K. Kazymyrenko and X. Waintal, Phys. Rev. B {\bf 77}, 115119 (2008).

%Fano
\bibitem{Tekman93} E. Tekman and P.~F. Bagwell, Phys. Rev. B {\bf 48}, 2553 (1993).
\bibitem{Kim99} C.~S. Kim, A.~M. Satanin, Y.~S. Joe and R.~M. Cosby, Phys. Rev. B {\bf 60}, 10962 (1999).
\bibitem{Vargiamidis05} V. Vargiamidis and H.~M. Polatoglou, Phys. Rev. B {\bf 72}, 195333 (2005).

\bibitem{noteunscreened} We do, indeed, get the same results as in Ref.~\onlinecite{Rurali08} for {\it unscreened} impurities: The transmission of minority carriers is blocked over a few hundreds of meV around the band edges. However, the transport of minority carriers mostly happen in gated devices, so that the impurities are, in practice, screened by oxides, gates and free carriers.

%Note 1
\bibitem{note1} The Coulomb potentials for P and B are not exactly equal in absolute value because they differ on the impurity site, corresponding to different chemical shifts.\cite{Diarra07,Diarra08}

\bibitem{notevalley} The $\Delta_2$ and $\Delta_4$ valleys are exactly two- and fourfold degenerate in the effective mass approximation only. In the tight-binding approach, intervalley couplings completely lift the degeneracies within the $\Delta_2$ and $\Delta_4$ valleys. This is for example visible in Fig. \ref{fig_comparison_conductance}: The {\it two} steps around $\varepsilon=1.33$ eV actually correspond to the $\Delta_4$ valleys.

\bibitem{Masetti83} G. Masetti, M. Severi, and S. Solmi, IEEE Trans. Dev. Electron Dev. {\bf 30}, 764 (1983).
\bibitem{Mole99} P.~J. Mole, J.~M. Rorison, J.~A. del Alamo and D. Lancefield, in {\it Properties of Crystalline Silicon}, ed. by Robert Hull (INSPEC, 1999).
\bibitem{Swirhun86} S.~E. Swirhun, Y.~H. Kwark, and R.~M. Swanson, IEDM Technical Digest 1986, 27 (1986).

%variability
\bibitem{Kuhn08} K.~J. Kuhn, C. Kenyon, A. Kornfeld, M. Liu, A. Maheshwari, W.-K. Shih, S. Sivakumar, G. Taylor, P. VanDerVoorn, and K. Zawadzki, Intel Technology Journal {\bf 12}, 93 (2008).

\end{thebibliography}
\end{document}